\documentclass[twocolumn,pra,showpacs]{revtex4}
\usepackage{amssymb}
\usepackage{graphicx}

\input{tcilatex}

\begin{document}

\title{Quantum-classical transition of the escape rate of uniaxial
antiferromagnetic particles in an arbitrarily directed field}
\author{Bin Zhou$^{1,2,3}$, Ruibao Tao$^{1}$, and Shun-Qing Shen$^{2}$}
\affiliation{$^{1}$Department of Physics, Fudan University, Shanghai 200433, China\\
$^{2}$Department of Physics, The University of Hong Kong, Hong Kong, China\\
$^{3}$Department of Physics, Hubei University, Wuhan 430062, China}
\date{\today}

\begin{abstract}
\thinspace Quantum-classical escape rate transition has been studied for
uniaxial antiferromagnetic particles with an arbitrarily directed magnetic
field. In the case that the transverse and longitudinal fileds coexist, we
calculate the phase boundary line between first- and second-order
transitions, from which phase diagrams can be obtained. It is shown that the
effects of the applied longitudinal magnetic field on quantum-classical
transition vary greatly for different relative magnitudes of the
non-compensation.
\end{abstract}

\pacs{75.50.Xx, 75.45.+j, 03.65.Sq}
\maketitle

Escape from a stable or metastable state at high temperatures is governed by
a classical thermal activation rate. At low temperatures close to zero,
quantum tunneling becomes relevant. When these two escape rates are equal
there exists a crossover temperature $T_{0}$ at which a transition between
classical and quantum regimes occurs. The study of the quantum-classical
transition is an interesting subject with a long history.\cite{Affleck} One
of the main issues in this subject is to determine whether the transition is
first- or second-order. The transition was recognized as a smooth
second-order one in the quantum mechanical models of Affleck\cite{Affleck}
and the cosmological models of Linde.\cite{Linde} However, it was shown\cite%
{Chudnovsky} that the smooth transition is not generic. Chudnovsky has
suggested that the order of transition is determined by the behavior of the
Euclidean time oscillation period $\tau (E)$, where $E$ is the energy near
the bottom of the Euclidean potential, which corresponds to the top of the
potential barrier.\cite{Chudnovsky} The non-monotonic behavior of the
oscillation period as a function of energy, i.e., the existence of a minimum
in the $\tau \sim E$ curve, was proposed as a condition for the first order
transition in quantum-mechanical tunneling.\cite{Chudnovsky} Later, a
sufficient criterion for the first-order phase transition was obtained by
carrying out the nonlinear perturbation near the sphaleron solution.\cite%
{Gorokhov}

Since the first- and second-order transitions between the quantum and
classical behaviors of the escape rates in spin systems were introduced by
Chudnovsky and Garanin,\cite{Chudnovsky (b), Garanin} the topic has
attracted considerate attention.\cite{Muller, Liang (b), Kim(a), Kim,
Garanin(a), Garanin(b), Martinez, Zhou(a), Lee (a)} Most theoretical studies
have been focused on the ferromagnetic particles. However, most
ferromagnetic systems are actually ferrimagnetic particles. For instance,
both Mn$_{12}$Ac and Fe$_{8}$ are characterized by a large spin ground state
which originates from incomplete compensation of antiferromagnetically
coupled spins.\cite{Caneschi} The strong exchange interaction should be
taken into consideration. In Ref.\cite{Kim(b)} Kim treated the phase
transition in ferrimagnetic or antiferromagnetic particles for two general
forms of the magnetic anisotropy energy. Very recently, the
quantum-classical transition in antiferromagnetic particles with biaxial
symmetry in the presence of an applied magnetic field along the medium axis
or along the easy axis was investigated.\cite{Zhou, Kim(c)} Note that recent
work of Chudnovsky and Garanin postulates dislocations as the main source of
spin tunneling in Mn$_{12}$ crystals.\cite{Chudnovsky(c)} Their theory shows
that when the external magnetic field is applied along the $c$ axis of the
crystal, local rotations of the magnetic anisotropy axis due to dislocations
result in the effective local transverse magnetic field. Experimental
evidence of the effects of dislocations on tunneling has been also reported.%
\cite{Park,Parks,Mertes,Hernandez} Therefore in the study of the
quantum-classical transition of Mn$_{12}$, the case of coexistence of the
transverse and longitudinal magnetic field is worth investigating.
Considering the molecular cluster Mn$_{12}$ actually is ferrimagnetic, the
exchange interaction should be also taken into account. In this paper we aim
to investigate the quantum-classical transition of the escape rate of
uniaxial antiferromagnetic particles in an arbitrarily directed field, i.e.,
the coexistence of the transverse and longitudinal magnetic field. It is
shown that the effects of the applied longitudinal magnetic field on the
quantum-classical transition vary greatly for different relative magnitudes
of the non-compensation.

We consider a small uniaxial antiferromagnetic particle with two magnetic
sublattices whose magnetizations, $\mathbf{m}_{1}$ and $\mathbf{m}_{2}$, are
coupled by the strong exchange interaction $\mathbf{m}_{1}\cdot \mathbf{m}%
_{2}/\chi _{\bot }$, where $\chi _{\bot }$ is the perpendicular
susceptibility. The system of interest has a non-compensation of sublattice
with $m$ ($=m_{1}-m_{2}>0$), and easy-axis anisotropy along the $z$-axis. In
the presence of an arbitrarily directed magnetic field, i.e., the
coexistence of a transverse magnetic field $H_{x}$ along the $x$-axis and a
longitudinal one $H_{z}$ along the $z$-axis, the Euclidean action is written
as\cite{Chudnovsky(d)}%
\begin{eqnarray}
S_{E}(\theta ,\phi ) &=&V\int d\tau \{i\frac{m_{1}+m_{2}}{\gamma }\dot{\phi}%
-i\frac{m}{\gamma }\dot{\phi}\cos \theta   \nonumber \\
&&+\frac{\widetilde{\chi }_{\bot }}{2\gamma ^{2}}[(\dot{\theta}+i\gamma
H_{x}\sin \phi )^{2}+(\dot{\phi}\sin \theta   \nonumber \\
&&+i\gamma H_{x}\cos \theta \cos \phi -i\gamma H_{z}\sin \theta )^{2}] 
\nonumber \\
&&+K_{\parallel }\sin ^{2}\theta -mH_{x}\sin \theta \cos \phi   \nonumber \\
&&-mH_{z}\cos \theta \},  \label{SE}
\end{eqnarray}%
where $V$ is the volume of the particle, $\gamma $ the gyromagnetic ratio, $%
\widetilde{\chi }_{\bot }=$ $\chi _{\bot }(m_{2}/m_{1})$ and $K_{\parallel }$
the longitudinal anisotropy. The polar coordinate $\theta $ and the
azimuthal coordinate $\phi $, which are the angular components of $\mathbf{m}%
_{1}$ in the spherical coordinate system, can determine the direction of the
N\'{e}el vector. Dot over a symbol denotes a derivative with respect to the
Euclidean time $\tau $.

The classical trajectory corresponding to the Euclidean action (1) is
determined by the equations 
\begin{eqnarray}
&&in\dot{\phi}\sin \theta +x(-ib_{z}\dot{\phi}\sin 2\theta -2ib_{x}\dot{\phi}%
\cos \phi \sin ^{2}\theta -\ddot{\theta}  \nonumber \\
&&+\frac{1}{2}\dot{\phi}^{2}\sin 2\theta )-\frac{V_{\theta }}{K_{\parallel }}
\nonumber \\
&=&0,  \label{CE1}
\end{eqnarray}%
\begin{eqnarray}
&&-in\dot{\theta}\sin \theta +x(ib_{z}\dot{\theta}\sin 2\theta +2ib_{x}\dot{%
\theta}\cos \phi \sin ^{2}\theta   \nonumber \\
&&-\ddot{\phi}\sin ^{2}\theta -\dot{\phi}\dot{\theta}\sin 2\theta )-\frac{%
V_{\phi }}{K_{\parallel }}  \nonumber \\
&=&0,  \label{CE2}
\end{eqnarray}%
where $n=m/(K_{\parallel }\gamma )$, $x=\widetilde{\chi }_{\bot
}/(K_{\parallel }\gamma ^{2})$, $b_{x(z)}=\gamma H_{x(z)}$, $V_{\theta
}=\partial V/\partial \theta $ and $V_{\phi }=\partial V/\partial \phi $.
The inverted potential is 
\begin{eqnarray}
V(\theta ,\phi ) &=&K_{\parallel }(-\sin ^{2}\theta +2h_{x}\sin \theta \cos
\phi +2h_{z}\cos \theta   \nonumber \\
&&+\frac{xb_{x}^{2}}{2}\sin ^{2}\phi +\frac{xb_{x}^{2}}{2}\cos ^{2}\theta
\cos ^{2}\phi   \nonumber \\
&&+\frac{xb_{z}^{2}}{2}\sin ^{2}\theta -xb_{x}b_{z}\sin \theta \cos \theta
\cos \phi ),
\end{eqnarray}%
where $h_{x(z)}=H_{x(z)}/H_{c}$ and $H_{c}=2K_{\parallel }/m$. In Fig.1 the
effective potential $-V(\theta ,\phi =0)$ is drawn. The minima of the
potential correspond to the equilibrium orientation of the N\'{e}el vector.
The metastability condition that $dV(\theta ,0)/d\theta =0$ and $%
d^{2}V(\theta ,0)/d\theta ^{2}=0$ determines the critical parameters at
which the barrier vanishes.\FRAME{ftbpFU}{3.2491in}{2.2191in}{0pt}{\Qcb{ The
effective potential $-V(\protect\theta ,\protect\phi =0).$}}{}{figure1.eps}{%
\special{language "Scientific Word";type "GRAPHIC";display
"USEDEF";valid_file "F";width 3.2491in;height 2.2191in;depth
0pt;original-width 3.3615in;original-height 3.1358in;cropleft "0";croptop
"1";cropright "1";cropbottom "0";filename 'figure1.eps';file-properties
"XNPEU";}}

\FRAME{ftbpFU}{3.2802in}{2.2399in}{0pt}{\Qcb{ $\protect\theta _{0}$ versus $%
h_{z}$ for some given $h_{x}$ and $y.$}}{}{figure2.eps}{\special{language
"Scientific Word";type "GRAPHIC";display "USEDEF";valid_file "F";width
3.2802in;height 2.2399in;depth 0pt;original-width 3.3615in;original-height
3.1358in;cropleft "0";croptop "1";cropright "1";cropbottom "0";filename
'figure2.eps';file-properties "XNPEU";}}

In the high-temperature regime the sphaleron solution of Eqs.(\ref{CE1}) and
(\ref{CE2}) is $(\theta _{0},\phi _{0}=0)$. $\theta _{0}$ is the position of
the top of the potential barrier $-V(\theta _{0},0)$, and is determined by $%
dV(\theta ,0)/d\theta |_{\theta _{0}}=0$ and $d^{2}V(\theta ,0)/d\theta
^{2}|_{\theta _{0}}>0$. $\theta _{0}$ has a cumbersome analytical form and
its numerical result will be applied below to determine phase boundary
lines. Furthermore, its behavior is illustrated in Fig.2 for given $h_{x}$
and $y$. Above, $y=x/n^{2}(=\tilde{\chi}_{\bot }K_{\parallel }/m^{2})$ and
the parameter $y$ indicates the relative magnitude of the non-compensation.
For large non-compensation ($y\ll 1,$ i.e., $m\gg \sqrt{\tilde{\chi}_{\bot
}K_{\parallel }}$) and for small non-compensation ($y\gg 1,$ i.e., $m\ll 
\sqrt{\tilde{\chi}_{\bot }K_{\parallel }}$), the system becomes
ferromagnetic and nearly compensated antiferromagnetic, respectively.\cite%
{Kim(b)} Note that $\theta _{0}=\pi /2$ for $h_{z}=0$. The crossover
behavior of the escape rate of this model from quantum tunneling to thermal
activation can be obtained from the deviation of the period of the periodic
instanton from that of the sphaleron. To this end we expand $(\theta ,\phi )$
about the sphaleron configurations $\theta _{0}$ and $\phi _{0}$, i.e., $%
\theta =\theta _{0}+\eta (\tau )$ and $\phi =\phi _{0}+\xi (\tau )$, where $%
\phi _{0}=0$. Substituting them into Eqs.(\ref{CE1}) and (\ref{CE2}) one
yields the following power series equations of the fluctuation fields $\eta $
and $\xi $%
\begin{equation}
\left( 
\begin{array}{c}
G_{1}^{\xi }(\eta ,\xi ) \\ 
G_{1}^{\eta }(\eta ,\xi )%
\end{array}%
\right) +\left( 
\begin{array}{c}
G_{2}^{\xi }(\eta ,\xi ) \\ 
G_{2}^{\eta }(\eta ,\xi )%
\end{array}%
\right) +\left( 
\begin{array}{c}
G_{3}^{\xi }(\eta ,\xi ) \\ 
G_{3}^{\eta }(\eta ,\xi )%
\end{array}%
\right) +...=0,  \label{FE}
\end{equation}%
where $G_{1},G_{2},G_{3},...$denote terms which contain linear, quadratic,
cubic and higher powers of the small fluctuations, respectively:%
\begin{eqnarray*}
G_{1}^{\xi }(\eta ,\xi ) &=&in\sin \theta _{0}\dot{\xi}-x\ddot{\eta}%
+A_{1}\eta  \\
&&-ix(b_{z}\sin 2\theta _{0}+2b_{x}\sin ^{2}\theta _{0})\dot{\xi},
\end{eqnarray*}%
\begin{eqnarray*}
G_{2}^{\xi }(\eta ,\xi ) &=&in\cos \theta _{0}\eta \dot{\xi}+\frac{1}{2}%
x\sin 2\theta _{0}\dot{\xi}^{2}+A_{2}\xi ^{2}+A_{4}\eta ^{2} \\
&&-2ix(b_{x}\sin 2\theta _{0}+b_{z}\cos 2\theta _{0})\eta \dot{\xi},
\end{eqnarray*}%
\begin{eqnarray*}
G_{3}^{\xi }(\eta ,\xi ) &=&-\frac{i}{2}n\sin \theta _{0}\eta ^{2}\dot{\xi}%
+x\cos 2\theta _{0}\eta \dot{\xi}^{2}+A_{3}\eta \xi ^{2}+A_{5}\eta ^{3} \\
&&+2ix(b_{z}\sin 2\theta _{0}-b_{x}\cos 2\theta _{0})\eta ^{2}\dot{\xi} \\
&&+ixb_{x}\sin ^{2}\theta _{0}\xi ^{2}\dot{\xi},
\end{eqnarray*}%
\begin{eqnarray*}
G_{1}^{\eta }(\eta ,\xi ) &=&in\sin \theta _{0}\dot{\eta}-ix(b_{z}\sin
2\theta _{0}+2b_{x}\sin ^{2}\theta _{0})\dot{\eta} \\
&&+x\sin ^{2}\theta _{0}\ddot{\xi}+B_{1}\xi ,
\end{eqnarray*}%
\begin{eqnarray*}
G_{2}^{\eta }(\eta ,\xi ) &=&in\cos \theta _{0}\eta \dot{\eta}-2ix(b_{x}\sin
2\theta _{0}+b_{z}\cos 2\theta _{0})\eta \dot{\eta} \\
&&+x\sin 2\theta _{0}(\dot{\eta}\dot{\xi}+\eta \ddot{\xi})+B_{2}\eta \xi ,
\end{eqnarray*}%
\begin{eqnarray*}
G_{3}^{\eta }(\eta ,\xi ) &=&-\frac{1}{2}in\sin \theta _{0}\eta ^{2}\dot{\eta%
}+ixb_{x}\sin ^{2}\theta _{0}\xi ^{2}\dot{\eta} \\
&&+2ix(b_{z}\sin 2\theta _{0}-b_{x}\cos 2\theta _{0})\eta ^{2}\dot{\eta} \\
&&+x\cos 2\theta _{0}(2\eta \dot{\eta}\dot{\xi}+\eta ^{2}\ddot{\xi}%
)+B_{3}\eta ^{2}\xi +B_{4}\xi ^{3},
\end{eqnarray*}%
where%
\begin{eqnarray}
A_{1} &=&-\frac{V_{\theta \theta }}{K_{\parallel }},A_{2}=-\frac{V_{\theta
\phi \phi }}{2K_{\parallel }},A_{3}=-\frac{V_{\theta \theta \phi \phi }}{%
2K_{\parallel }},  \nonumber \\
A_{4} &=&-\frac{V_{\theta \theta \theta }}{2K_{\parallel }},A_{5}=-\frac{%
V_{\theta \theta \theta \theta }}{6K_{\parallel }},
\end{eqnarray}%
\begin{equation}
B_{1}=\frac{V_{\phi \phi }}{K_{\parallel }},B_{2}=\frac{V_{\theta \phi \phi }%
}{K_{\parallel }},B_{3}=\frac{V_{\theta \theta \phi \phi }}{2K_{\parallel }}%
,B_{4}=\frac{V_{\phi \phi \phi \phi }}{6K_{\parallel }}.
\end{equation}%
\ It is introduced that $V_{\theta \theta }=[\partial ^{2}V/\partial \theta
^{2}]_{\theta =\theta _{0},\phi =\phi _{0}}$, $V_{\theta \phi \phi
}=[\partial ^{2}V/\partial \theta \partial \phi ^{2}]_{\theta =\theta
_{0},\phi =\phi _{0}}$, and so on.

Denoting $\delta \Omega (\tau )\equiv (\eta (\tau ),\xi (\tau ))$, we have $%
\delta \Omega (\tau +\beta \hbar )=$ $\delta \Omega (\tau )$ at finite
temperature and write it as the Fourier series $\delta \Omega (\tau
)=\sum_{n=-\infty }^{\infty }$ $\delta \Omega _{n}\exp [i\omega _{n}\tau ]$,
where $\omega _{n}=2\pi n/\beta \hbar $. Since simple analysis shows that $%
\eta $ is real and $\xi $ imaginary, to the lowest order we write them in
the form $\eta \simeq a\theta _{1}\cos (\omega \tau )$ and $\xi \simeq
ia\phi _{1}\sin (\omega \tau )$. Here $a$ serves as a perturbation
parameter. Substituting them into Eq.(\ref{FE}) and neglecting terms of
order higher than $a$, we obtain the relation%
\begin{eqnarray}
\frac{\phi _{1}}{\theta _{1}} &=&\frac{x\omega _{\pm }^{2}+A_{1}}{\omega
_{\pm }[n-2x(b_{z}\cos \theta _{0}+b_{x}\sin \theta _{0})]\sin \theta _{0}} 
\nonumber \\
&=&-\frac{\omega _{\pm }[n-2x(b_{z}\cos \theta _{0}+b_{x}\sin \theta
_{0})]\sin \theta _{0}}{x\omega _{\pm }^{2}\sin ^{2}\theta _{0}-B_{1}}
\end{eqnarray}%
and the oscillation frequency%
\begin{eqnarray}
\omega _{\pm }^{2} &=&-\frac{1}{2x^{2}}\{(A_{1}-B_{1}\csc ^{2}\theta _{0})x 
\nonumber \\
&&+[n-2x(b_{z}\cos \theta _{0}+b_{x}\sin \theta _{0})]^{2}\}  \nonumber \\
&&\pm \frac{1}{2x^{2}}(4A_{1}B_{1}x^{2}\csc ^{2}\theta
_{0}+\{(A_{1}-B_{1}\csc ^{2}\theta _{0})x  \nonumber \\
&&+[n-2x(b_{z}\cos \theta _{0}+b_{x}\sin \theta _{0})]^{2}\}^{2})^{1/2}.
\end{eqnarray}

Next, let us write $\eta \simeq a\theta _{1}\cos (\omega \tau )+\eta _{2}$,
and $\xi \simeq ia\phi _{1}\sin (\omega \tau )+i\xi _{2}$, where $\eta _{2}$
and $\xi _{2}$ are of the order of $a^{2}$. Inserting them into Eq.(\ref{FE}%
), we arrive at $\omega =\omega _{+}$ and 
\begin{equation}
\eta _{2}=a^{2}p_{0}+a^{2}p_{2}\cos (2\omega \tau ),\text{ \ }\xi
_{2}=a^{2}q_{2}\sin (2\omega \tau ),  \label{second}
\end{equation}%
where the analytic forms of coefficients $p_{0}$, $p_{2}$ and $q_{2}$ are
cumbersome, which are listed in the Appendix.

This implies that there is no shift in the oscillation frequency. In order
to find the change of the oscillation period, we proceed to the third order
of perturbation theory by writing $\eta \simeq a\theta _{1}\cos (\omega \tau
)+\eta _{2}+\eta _{3}$, and $\xi \simeq ia\phi _{1}\sin (\omega \tau )+i\xi
_{2}+i\xi _{3}$, where $\eta _{3}$ and $\xi _{3}$ are of the order of $a^{3}$%
. Substituting them again into Eq.(\ref{FE}), and retaining only the terms
up to $O(a^{3})$, we have 
\begin{equation}
n^{4}y^{2}(\omega ^{2}-\omega _{+}^{2})(\omega ^{2}-\omega _{-}^{2})=a^{2}%
\frac{\theta _{1}^{2}}{4\sin ^{2}\theta _{0}}g(h_{x},h_{z},y),
\label{result}
\end{equation}%
where 
\begin{equation}
g(h_{x},h_{z},y)=g_{1}(h_{x},h_{z},y)+g_{2}(h_{x},h_{z},y).  \label{G}
\end{equation}%
The forms of $g_{1}(h_{x},h_{z},y)$ and $g_{2}(h_{x},h_{z},y)$ are 
\begin{eqnarray}
g_{1}(h_{x},h_{z},y) &=&-(A_{1}+yw^{2})\{3B_{4}\lambda ^{2}-6\lambda
h_{x}w\sin ^{2}\theta _{0}  \nonumber \\
&&-B_{3}-2B_{2}(2\tilde{p}_{0}-\tilde{p}_{2})+3yw^{2}\cos 2\theta _{0} 
\nonumber \\
&&+2yw^{2}\sin 2\theta _{0}(2\tilde{p}_{0}+\tilde{p}_{2})+\frac{1}{\lambda }%
[-2B_{2}\tilde{q}_{2}  \nonumber \\
&&+2w\cos \theta _{0}(2\tilde{p}_{0}+\tilde{p}_{2})-4h_{x}yw\cos 2\theta _{0}
\nonumber \\
&&-\frac{1}{2}w\sin \theta _{0}-8h_{z}yw(2\tilde{p}_{0}+\tilde{p}_{2})\cos
2\theta _{0}  \nonumber \\
&&+4yw\sin 2\theta _{0}(h_{z}-4h_{x}\tilde{p}_{0}-2h_{x}\tilde{p}_{2}) 
\nonumber \\
&&+4\tilde{q}_{2}yw^{2}\sin 2\theta _{0}]\},
\end{eqnarray}%
\begin{eqnarray}
g_{2}(h_{x},h_{z},y) &=&(B_{1}-yw^{2}\sin ^{2}\theta _{0})\{2\lambda
^{3}h_{x}yw\sin ^{2}\theta _{0}  \nonumber \\
&&-\lambda ^{2}(A_{3}+3yw^{2}\cos 2\theta _{0})  \nonumber \\
&&+\lambda \lbrack -4A_{2}\tilde{q}_{2}-2(2\tilde{p}_{0}+\tilde{p}_{2})w\cos
\theta _{0}  \nonumber \\
&&+\frac{3}{2}w\sin \theta _{0}-4\tilde{q}_{2}yw^{2}\sin 2\theta _{0} 
\nonumber \\
&&+4yw\cos 2\theta _{0}(3h_{x}+4h_{z}\tilde{p}_{0}+2h_{z}\tilde{p}_{2}) 
\nonumber \\
&&-4yw\sin 2\theta _{0}(3h_{z}-4h_{x}\tilde{p}_{0}  \nonumber \\
&&-2h_{x}\tilde{p}_{2})]+4A_{4}(2\tilde{p}_{0}+\tilde{p}_{2})  \nonumber \\
&&+4\tilde{q}_{2}w(4h_{z}y\cos 2\theta _{0}+4h_{x}y\sin 2\theta _{0} 
\nonumber \\
&&-\cos \theta _{0})+3A_{5}\},
\end{eqnarray}%
where $w=n\omega _{+}$ and $\lambda =\phi _{1}/\theta _{1}$. Again, $%
y=x/n^{2}(=\tilde{\chi}_{\bot }K_{\parallel }/m^{2})$ and the parameter $y$
indicates the relative magnitude of the non-compensation. Also, $\tilde{p}%
_{0}$, $\tilde{p}_{2}$ and $\tilde{q}_{2}$ are obtained by replacing $\phi
_{1}$ by $\lambda \theta _{1}$ and dropping $\theta _{1}^{2}$ in $p_{0}$, $%
p_{2}$ and $q_{2}$, respectively. It can be shown that for $h_{z}=0$ Eq.(\ref%
{G}) is reduced to the case corresponding to uniaxial antiferromagnetic
particle with a transverse magnetic field only has been investigated in Ref.%
\cite{Kim(b)}.\FRAME{ftbpFU}{3.1765in}{2.1681in}{0pt}{\Qcb{Phase diagram $%
h_{x}$($y)$ for $h_{z}=0,0.1$ and $0.2$.}}{}{figure3.eps}{\special{language
"Scientific Word";type "GRAPHIC";display "USEDEF";valid_file "F";width
3.1765in;height 2.1681in;depth 0pt;original-width 3.3615in;original-height
3.1358in;cropleft "0";croptop "1";cropright "1";cropbottom "0";filename
'figure3.eps';file-properties "XNPEU";}}

\FRAME{ftbpFU}{3.1765in}{2.1681in}{0pt}{\Qcb{Phase diagram $h_{x}$($h_{z})$
for $y=0$ and $0.1$.}}{}{figure4.eps}{\special{language "Scientific
Word";type "GRAPHIC";display "USEDEF";valid_file "F";width 3.1765in;height
2.1681in;depth 0pt;original-width 3.3615in;original-height 3.1358in;cropleft
"0";croptop "1";cropright "1";cropbottom "0";filename
'figure4.eps';file-properties "XNPEU";}}

\FRAME{ftbpFU}{3.1765in}{2.1681in}{0pt}{\Qcb{Phase diagram $h_{x}$($h_{z})$
for $y=0.2$ and $0.3$.}}{}{figure5.eps}{\special{language "Scientific
Word";type "GRAPHIC";display "USEDEF";valid_file "F";width 3.1765in;height
2.1681in;depth 0pt;original-width 3.3615in;original-height 3.1358in;cropleft
"0";croptop "1";cropright "1";cropbottom "0";filename
'figure5.eps';file-properties "XNPEU";}}

As shown by Chudnovsky\cite{Chudnovsky}, if the oscillation period $\tau $
is not a monotonic function of $a$, where $a$ is a function of $E$ in the
absence of dissipation, the system exhibits a first-order transition. Thus,
the period $\tau (=2\pi /\omega )$ in Eq.(\ref{result}) should be less than $%
\tau _{+}(=2\pi /\omega _{+})$, i.e., $\omega >\omega _{+}$ for the
first-order transition. It implies that $g(h_{x},h_{z},y)>0$ in Eq.(\ref%
{result}) for the first-order transition, and $g(h_{x},h_{z},y)=0$
determines the phase boundary between the first- and the second-order
transition. In this case the three parameters $h_{x},h_{z},y$ should be
treated simultaneously, which is not a simple problem. In the present work
we will fix one parameter and then compute the boundary curve with the other
two parameters. We first solve Eq.(\ref{result}) numerically to obtain the
phase boundary lines$\ h_{x}(y)$'s for several values of $h_{z}$, which are
plotted in Fig.3. From Fig.3, an immediate observation is that the
first-order region for a given $h_{z}$ diminishes as $y$ increases, which
shows the same trend as the $h_{z}=0$ case. Thus, it is evident that the
region for the first-order transition is greatly reduced as the system
becomes ferrimagnetic and there is no first-order transition in almost
compensated antiferromagnetic particles. The result coincides with Ref.\cite%
{Kim(b)}. Fig.3 also shows that with increasing $h_{z}$ the variety of
first-order region is not a simple case. For the small $y$ case with
increasing $h_{z}$ the first-order region is shrunk, while for the larger $y$
case the longitudinal field $h_{z}$ favors occurrence of the first-order
transition. For instance, for the case of $y=0.05$, the maximum values of
the transverse filed $h_{x}$ for occurrence of the first-order transition
are $h_{x}=0.203,$ $0.195$ and $0.175$ for $h_{z}=0,$ $0.1$ and $0.2$,
respectively. On the other hand, the first-order region vanishes beyond $%
y\simeq 0.46,$ $0.55$ and $0.89$ for $h_{z}=0,$ $0.1$ and $0.2$,
respectively. This can be qualitatively understood from the consideration
that the height of the effective potential barrier decreases as $h_{z}$
increases, whereas the height increases as $y$ increases, therefore there is
a competition between the longitudinal field and the relative magnitude of
the non-compensation. When $y=0$, the fact that the region for the
first-order transition decreases as the longitudinal field increases results
from a flattening of the peak of the barrier.\cite{Garanin(a)} For the small 
$y$ case (i.e., the large noncompensation), the crossover behavior of the
ferrimagnetic system still keeps qualitatively that of the ferromagnetic
one. However, for the larger $y$ case the exchange interaction plays the
role of effective magnetic field and so, for a given small transverse field
the region for the first-order transition increases as the longitudinal
field increases. To illustrate further the effect of the longitudinal field $%
h_{z}$ on quantum-classical transition, we next calculate the phase boundary
lines$\ h_{x}(h_{z})$'s for several values of $y$, which are shown in Figs.4
and 5. In Fig.4, for the case of $y=0$ corresponding to uniaxial
ferromagnetic system, the phase boundary line is plotted by a dotted line.
Obviously the line coincides with Fig.13 in Ref.\cite{Garanin(a)}, in which
quantum-classical transition in a uniaxial ferromagnetic system with a
transverse magnetic field and a longitudinal one was investigated. For the
case of $y=0.1$, the phase boundary lines$\ h_{x}(h_{z})$'s shift downwards
and with $h_{z}$ increasing the critical value of $h_{x}$ decrease
monotonically. Fig.5 gives another case, in which the phase boundary lines$\
h_{x}(h_{z})$'s show a kind of non-monotonic behavior. For instance, for the
case of $y=0.3$, the first-order region vanishes beyond $h_{x}\simeq 0.024$
for $h_{z}=0$, while the maximum is $h_{x}\simeq 0.112$ for $h_{z}=0.332$.

It was shown that quantum tunneling shall show up at higher temperatures and
higher frequencies in antiferromagnetic particles than in ferromagnetic
particles of similar size.\cite{Chudnovsky(d)} Moreover, most ferromagnetic
systems are ferrimagnetic, so nanometer-scale antiferromagnets are more
interesting from experimental and theoretical aspects. But a detailed
comparison between the theory and experiment on quantum-classical transition
remains a challenging task. It is very important to obtain the information
on the magnitude of the quantity $y$ for observing the first-order
transition in real experiments. For the typical antiferromagnetic particle
with $\chi \sim 10^{-4}$, $K_{\parallel }\sim 10^{6}$ erg/cm$^{3}$, and $%
m\sim 500$ emu/cm$^{3}$, one can get the quantity $y\simeq 10^{-4}$.\cite%
{Kim(c)} In this case, for the longitudinal field parameter $h_{z}=0.4$, the
range of the transverse field parameter for observing the first-order
transition is $0<h_{x}\lesssim 0.127$. It is note that Wensdorfer et al have
performed the switching field measurements on individual ferrimagnetic and
insulating BaFeCoTiO nano-particle containing about $10^{5}$--$10^{6}$ spins
at very low temperature (0.1--6 K).\cite{Wernsdorfer} Below 0.4 K,
experimental results are quantitatively in agreement with the predictions of
the macroscopic quantum tunneling theory without dissipation. The BaFeCoTiO
nano-particles have a strong uniaxial magnetocrystalline anisotropy.\cite%
{Wernsdorfer} Therefore, the material is expected as a candidate to
investigate quantum-classical transition of the escape rate of uniaxial
ferrimagnetic or antiferromagnetic particles in an arbitrarily directed
field.

In conclusion, we have investigated quantum-classical escape rate transition
for uniaxial antiferromagnetic particle with an arbitrarily directed
magnetic field, i.e., the coexistence of the transverse and longitudinal
magnetic fields. There are three parameters which can be controlled by
experiment: relative magnitude of the non-compensation and two field
parameters. The nonlinear perturbation method is used to obtain various
phase diagrams for first- and second-order transition depending on the three
parameters. It is shown that the effects of the applied longitudinal
magnetic field on quantum-classical transition vary greatly for different
relative magnitudes of the non-compensation.\newline

The work was supported by China Postdoctoral Science Foundation under Grant
No. 2002032138, National Natural Science Foundation of China under Grant
Nos. 10174015 and 10234010, and Research Grant Council of Hong Kong.

\section*{Appendix}

The coefficients in Eq.(\ref{second}) are deduced by using software
Mathematica3.0.%
\begin{eqnarray*}
p_{0} &=&\frac{1}{2A_{1}}[\phi _{1}\cos \theta _{0}(n\theta _{1}\omega
+x\phi _{1}\omega _{+}^{2}\sin \theta _{0}) \\
&&+A_{2}\phi _{1}^{2}-A_{4}\theta _{1}^{2}-2x\omega \theta _{1}\phi
_{1}(b_{z}\cos 2\theta _{0} \\
&&+b_{x}\sin 2\theta _{0})],
\end{eqnarray*}%
\begin{eqnarray*}
p_{2} &=&\{B_{1}(A_{4}\theta _{1}^{2}+A_{2}\phi _{1}^{2})+2B_{2}n\omega
\theta _{1}\phi _{1}\sin \theta _{0} \\
&&-8b_{z}^{2}x^{2}\theta _{1}^{2}\omega ^{2}\cos ^{3}\theta _{0}\sin \theta
_{0} \\
&&-2x\omega (2B_{2}b_{x}\theta _{1}\phi _{1} \\
&&+B_{1}b_{z}\theta _{1}\phi _{1}+2A_{4}\theta _{1}^{2}\omega +2A_{2}\phi
_{1}^{2}\omega )\sin ^{2}\theta _{0} \\
&&+8b_{z}x^{2}\theta _{1}\omega ^{2}(b_{x}\theta _{1}+\omega \phi _{1})\sin
^{4}\theta _{0} \\
&&-4b_{z}nx\theta _{1}^{2}\omega ^{2}\sin ^{3}\theta _{0}+2b_{z}x\theta
_{1}\omega \cos ^{2}\theta _{0}[B_{1}\phi _{1} \\
&&-6b_{x}x\theta _{1}\omega -2x\phi _{1}\omega ^{2}+4x\phi _{1}\omega
_{+}^{2} \\
&&+2x(3b_{x}\theta _{1}\omega +\phi _{1}\omega ^{2}-2\phi _{1}\omega
_{+}^{2})\cos 2\theta _{0} \\
&&+4n\theta _{1}\omega \sin \theta _{0}]-\cos \theta _{0}[B_{1}n\theta
_{1}\phi _{1}\omega  \\
&&+[2\theta _{1}\omega (n^{2}\theta _{1}\omega +2B_{2}b_{z}x\phi _{1}) \\
&&-B_{1}x\phi _{1}(4b_{x}\theta _{1}\omega -\phi _{1}\omega _{+}^{2})]\sin
\theta _{0} \\
&&-4nx\theta _{1}\omega \lbrack 3b_{x}\theta _{1}\omega +\phi _{1}(\omega
^{2}-2\omega _{+}^{2})]\sin ^{2}\theta _{0} \\
&&+4x^{2}\omega \lbrack 4b_{x}^{2}\theta _{1}^{2}\omega +4b_{x}\theta
_{1}\phi _{1}(\omega ^{2}-\omega _{+}^{2}) \\
&&-\omega (2b_{z}^{2}\theta _{1}^{2}+\phi _{1}^{2}\omega _{+}^{2})]\sin
^{3}\theta _{0}]\}/\{-2B_{1}(A_{1}+4x\omega ^{2}) \\
&&+8\omega ^{2}[A_{1}x+(n-2b_{z}x\cos \theta _{0})^{2} \\
&&+4x^{2}\omega ^{2}]\sin ^{2}\theta _{0}+32b_{x}x\omega ^{2}(2b_{z}x\cos
\theta _{0}-n)\sin ^{3}\theta _{0} \\
&&+32b_{x}^{2}x^{2}\omega ^{2}\sin ^{4}\theta _{0}\},
\end{eqnarray*}

\begin{eqnarray*}
q_{2} &=&\{\theta _{1}(A_{1}+4x\omega ^{2})[B_{2}\phi _{1}-n\theta
_{1}\omega \cos \theta _{0} \\
&&+2x\theta _{1}\omega (b_{x}\sin 2\theta _{0}+b_{z}\cos 2\theta _{0}) \\
&&-2x\phi _{1}\omega _{+}^{2}\sin 2\theta _{0}]-\omega \sin \theta _{0}[n- \\
&&2x(b_{z}\cos \theta _{0}+b_{x}\sin \theta _{0})][2n\theta _{1}\phi
_{1}\omega \cos \theta _{0} \\
&&-4x\theta _{1}\phi _{1}\omega (b_{x}\sin 2\theta _{0}+b_{z}\cos 2\theta
_{0}) \\
&&+x\phi _{1}^{2}\omega _{+}^{2}\sin 2\theta _{0}-2A_{4}\theta
_{1}^{2}-2A_{2}\phi _{1}^{2}\} \\
&&/\{8\omega ^{2}\sin ^{2}\theta _{0}[n-2x(b_{z}\cos \theta _{0}+b_{x}\sin
\theta _{0})]^{2} \\
&&+2(A_{1}+4x\omega ^{2})(4x\omega ^{2}\sin ^{2}\theta _{0}-B_{1})\}.
\end{eqnarray*}

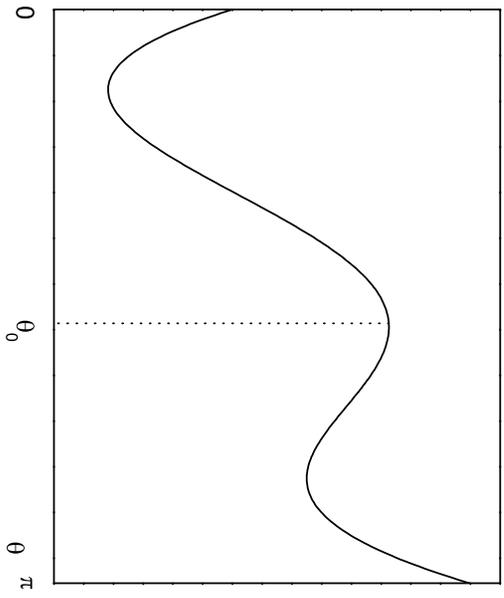
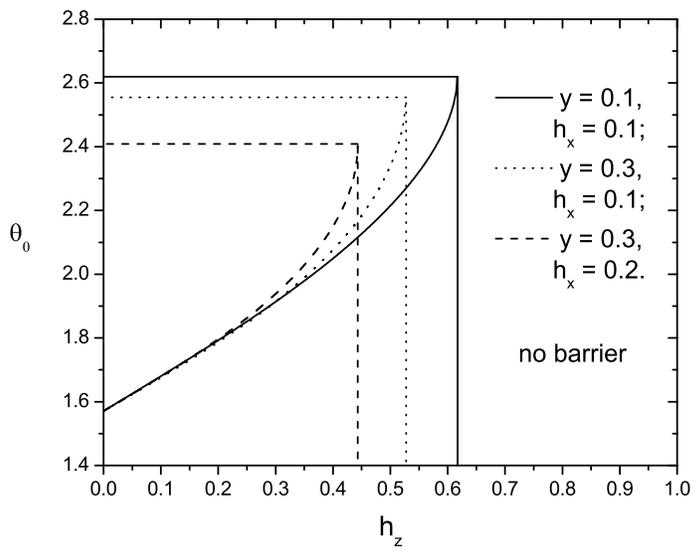
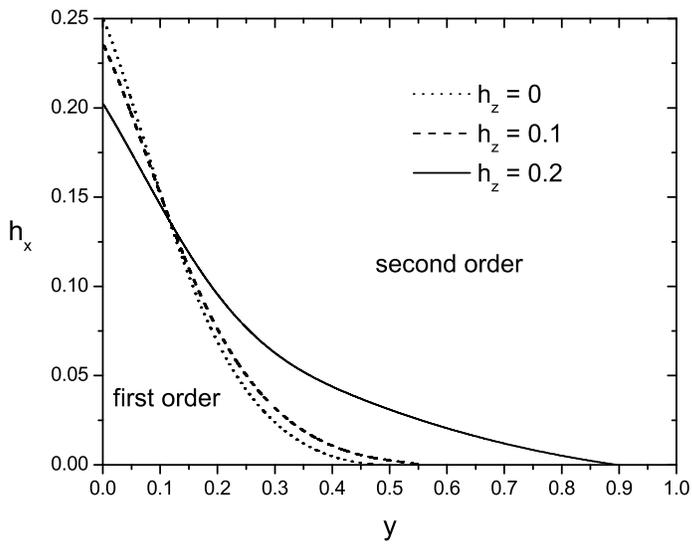
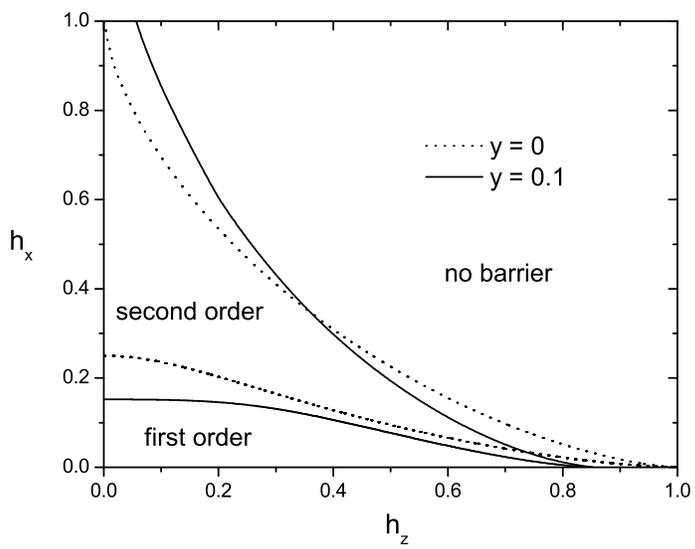
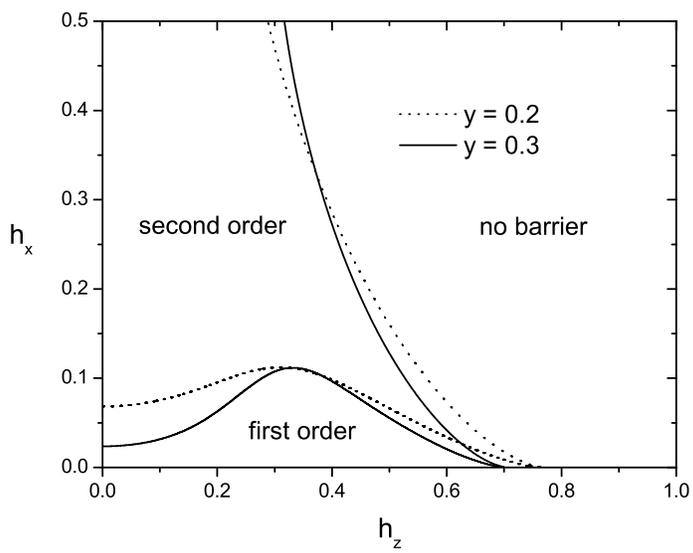
\end{document}